\RequirePackage{lineno}
\documentclass[floatfix,twocolumn,prd,showpacs,preprintnumbers,amsmath,nofootinbib,amssymb,superscriptaddress]{revtex4-1}

\usepackage{mathtools, cuted}
\usepackage{enumitem}
\usepackage{lipsum}
\usepackage[utf8]{inputenc}
\usepackage{graphicx}
\usepackage{dsfont}
\usepackage{mathrsfs}
\usepackage{bm}
\usepackage{color}
\usepackage[colorlinks,citecolor=blue,urlcolor=blue,linkcolor=black]{hyperref}
\usepackage{braket}
\usepackage{placeins}
\usepackage{multirow}
\usepackage{float}
\usepackage{booktabs}
\usepackage{soul}
\usepackage{float}
\usepackage{comment}

\usepackage[author={Luis}]{pdfcomment}

\usepackage{tikz,pgfplots,pgfplotstable}
\usepgfplotslibrary{groupplots,dateplot}
\usetikzlibrary{patterns,shapes.arrows}
\pgfplotsset{compat=newest}
\usepgfplotslibrary{fillbetween}
\usetikzlibrary{external}
\usepackage[nameinlink,capitalize]{cleveref}
\makeatletter
	\def\Ginput@path{{./figures/}}
\makeatother

\def\tf{\tau_{\text{F}}}

\def\OO{\mathcal{O}}

\def\Eq#1{Eq.~\eqref{#1}}
\def\scut{s_{\mathrm{cut}}}

\begin{document}

\title{Lattice QCD noise reduction for bosonic correlators through
blocking}

\author{Luis Altenkort}
\affiliation{Fakult\"at f\"ur Physik, Universit\"at Bielefeld, D-33615 Bielefeld, Germany}
\author{Alexander M.~Eller}
\affiliation{
    Institut f\"ur Kernphysik, Technische Universit\"at Darmstadt\\
Schlossgartenstra{\ss}e 2, D-64289 Darmstadt, Germany }
\author{O. Kaczmarek}
\affiliation{Fakult\"at f\"ur Physik, Universit\"at Bielefeld, D-33615 Bielefeld, Germany}

\author{Lukas Mazur}
\affiliation{Paderborn Center for Parallel Computing, Paderborn University, D-33098 Paderborn, Germany }
\author{Guy D.~Moore}
\affiliation{
    Institut f\"ur Kernphysik, Technische Universit\"at Darmstadt\\
Schlossgartenstra{\ss}e 2, D-64289 Darmstadt, Germany }
\author{Hai-Tao~Shu}
\affiliation{
    Institut f\"ur Theoretische Physik, Universit\"at   Regensburg, D-93040 Regensburg,
    Germany.
}

\begin{abstract}

We propose a method to substantially improve the signal-to-noise ratio of lattice correlation functions for bosonic operators
or other operator combinations with disconnected contributions.
The technique is applicable for correlations between operators on two planes (zero momentum correlators) when the dimension of the plane is larger than the separation between the two planes which are correlated.
In this case, the correlation arises primarily from points whose in-plane coordinates are close, but noise arises from all pairs of points.
By breaking each plane into bins and computing bin-bin correlations,
it is possible to capture these short-distance correlators exactly while replacing (small) correlators at large spatial extent with a fit, with smaller uncertainty than the data.
The cost is only marginally larger than averaging each plane before correlating, but the improvement in signal-to-noise can be substantial.
We test the method on correlators of the gradient-flowed topological charge density and squared field strength, finding noise reductions by a factor of $\sim$ 3$-$7
compared to the conventional approach on the same ensemble of configurations.
\end{abstract}

\maketitle
\section{Introduction}
\label{sec:intro}

Many problems in quantum field theory can be expressed in terms of the correlation function of two operators as a function of separation along one axis, while averaging over directions transverse to that axis.
For example, when the operators are the interpolating operator for a particle,
the exponential rate of the falloff of the correlator determines the particle mass. 
Moreover, zero spatial-momentum correlators of conserved currents, such as the energy-momentum tensor (EMT) or the vector current, encode transport coefficients like shear and bulk viscosity or flavor diffusion coefficients and the electrical conductivity in the small-frequency limiting behavior of their reconstructed spectral functions.
For a review see Ref.~\cite{Meyer:2011gj}. Some recent lattice studies using this approach for the calculation of viscosities can be found in Refs.~\cite{Astrakhantsev:2017nrs,Kitazawa:2017qab,Borsanyi:2018srz,Taniguchi:2019eid,Itou:2020azb}
and for a recent overview of results for the electrical conductivity see Ref.~\cite{Aarts:2020dda}.
Problems like these ultimately come down to the computation of correlation functions of operators averaged over a transverse plane,
evaluated as a function of the separation between two planes.

Usually such studies require that the correlation function be determined very precisely.  
For some operators built out of fermions which
carry nontrivial flavor, there are no disconnected contributions.
In this case, the precision is generally good, as
there are explicit factors of propagators between the two
planes, which cause the configuration-by-configuration value
of the correlator to decay.  In this case the signal to noise
is generally good,%
\footnote{In some cases, e.g, correlators of baryon operators, the signal-to-noise is bad even though there are only connected contributions \cite{Parisi:1983ae}.
This paper will not address this problem.}
and the main limitations are, for example, contamination
from higher states, the continuum limit, etc.
However, for correlation functions of operators consisting of bosonic fields, or correlators built out of fermions such that there are disconnected contributions, signal-to-noise problems are generically severe.
Recently there have been some advances in dealing with this problem,
such as, covariant coordinate-space methods \cite{Meyer:2017hjv, Ce:2018ziv}, frequency splitting methods \cite{Giusti:2019kff}, and the cluster decomposition method \cite{Liu:2017man}.

This paper will introduce and demonstrate a new approach for problems where such disconnected contributions lead to severe signal-to-noise problems.
In the next section, we review the origin of the noise in such disconnected correlation functions.
We show a numerically efficient way to express the correlation function between operators on two planes as an integral over the transverse separation between the operator locations on the planes in \autoref{sec:blocking_method}.
The signal is dominated by small transverse separations; the noise is dominated by large transverse separations.
By fitting the region with a strong signal and using the fit where the signal is poor, one can avoid these noisy contributions, improving the overall signal-to-noise significantly.
We demonstrate this for EMT correlators in the bulk channel and the topological charge density correlator in \autoref{sec:applications}. As a byproduct of this work we also update the double extrapolated topological charge density correlators which we provided recently \cite{Altenkort:2020axj}.

\section{Origin of the problem}
\label{sec:problem}

The origin of noise in correlators was first explored by Parisi \cite{Parisi:1983ae}, and our analysis follows his pioneering work.
Consider an operator $\OO$ of dimension $\Delta$ and its Hermitian conjugate $\OO^\dagger$.
Suppose we are interested in the correlation function
\begin{equation}
    G(\tau) \equiv \frac{1}{L^3}\int d^3\vec{x} d^3\vec{y} \;
    \langle \OO^\dagger(\vec{x},\tau) \OO(\vec{y},0) \rangle \,.
\end{equation}
Here the transverse space integration $d^3\vec{x}$ is over a transverse space of extent $L^3$, and we are primarily concerned with the case where $\tau \ll L$.
This is the case for transport coefficients because $\tau < 1/(2T)$ but $L \gg 1/T$ to ensure that we are close to the thermodynamic limit.
In the opposite limit, that is $\tau > L$, our approach will be ineffective.  For correlations between planes which include the time direction and are separated instead along a space axis, exchange the label $\tau$ for the label of the relevant space direction in what follows.

First let us analyze the expected size of the signal.
On dimensional grounds we expect that
\begin{equation}
\label{Gestimate}
\langle \OO^\dagger(\vec x, \tau) \OO(\vec y,0) \rangle \sim
 ((\vec x-\vec y)^2 + \tau^2)^{-\Delta} \,,
 \end{equation}
 \begin{align}
 G(\tau) & = \frac{1}{L^3}\int d^3 \vec x\ d^3 \vec y
 \langle \OO^\dagger(\vec x,\tau) \OO(\vec y,0) \rangle
 \nonumber \\
 & \propto \frac{1}{L^3}\int d^3 \frac{\vec x + \vec y}{2} \int d^3 (\vec x - \vec y)
 \frac{1}{((\vec x-\vec y)^2+\tau^2)^\Delta}
 \nonumber \\ &
 \sim \tau^{3-2\Delta} \,.
\end{align}
The correlation function is extensive in the transverse area because of the integral over the average coordinate $(\vec x + \vec y)/2$; the integral over the difference coordinate $\vec x - \vec y$ is dominated by $|\vec x - \vec y| \lesssim \tau$.

If a large mass gap plays a role in the correlator of interest, the decay will instead be exponential.
In general, one expects polynomial decay at short distances and exponential decay at large distances.

Next we want to understand the noise.
The signal-to-noise achieved from $N_{\mathrm{sample}}$ independent gauge field configurations will scale as $1/\sqrt{N_{\mathrm{sample}}}$ times the signal-to-noise from a single configuration.
We can estimate this noise by asking about the mean value of $G(\tau)$ and about the mean-squared value of $G(\tau)$.
Then the variance of the measurement is determined as usual by
\begin{widetext}
\begin{align}
\label{sigmaestimate}
\sigma_G^2 & = \langle G(\tau) G^*(\tau) \rangle
    - |\langle G(\tau) \rangle|^2
    \nonumber \\
    & = \frac{1}{L^6}\int d^3\vec{x_1}\, d^3\vec{x_2}\, d^3\vec{y_1}\, d^3\vec{y_2}
    \left( \langle \OO^\dagger(\vec{x_1},\tau) \OO(\vec{x_2},\tau)
    \OO(\vec{y_1},0) \OO^\dagger(\vec{y_2},0) \rangle
    - \langle \OO^\dagger(\vec{x_1},\tau) \OO(\vec{y_1},0) \rangle
    \langle \OO(\vec{x_2},\tau) \OO^\dagger(\vec{y_2},0) \rangle \right)
    \nonumber \\
    & \simeq \frac{1}{L^6}\int d^3\vec{x_1} d^3\vec{x_2} 
    \langle \OO^\dagger(\vec{x_1},\tau) \OO(\vec{x_2},\tau) \rangle
    \;\int d\vec{y_1} d\vec{y_2}
    \langle \OO(\vec{y_1},0) \OO^\dagger(\vec{y_2},0) \rangle 
    \nonumber \\
    & \sim \frac{1}{L^6}\int d^3 \frac{\vec x_1 + \vec x_2}{2} \int d^3 (\vec x_1 - \vec x_2)
    |\vec x_1 - \vec x_2|^{-2\Delta}
     \int d^3 \frac{\vec y_1 + \vec y_2}{2} \int d^3 (\vec y_1 - \vec y_2)
    |\vec y_1 - \vec y_2|^{-2\Delta} \,.
\end{align}
\end{widetext}
The first term in the second line is the $full$ correlator, including both the connected correlator and various disconnected contributions.
The second term cancels one of these disconnected contributions, but the disconnected contribution shown in the third line involves large small-separation contributions when $\vec x_1 \approx \vec x_2$ and
$\vec y_1 \approx \vec y_2$, and is therefore expected to dominate the correlation function.

The variance has two worrying features.
First, the integrals over $(\vec x_1-\vec x_2)$ and $(\vec y_1-\vec y_2)$ are short-distance divergent, presumably cut off by the lattice spacing $a$.
Second, $each$ overall integration $\int d^3 (\vec x_1+\vec x_2)/2$, $\int d^3 (\vec y_1+\vec y_2)/2$ introduces an overall $L^3$ factor.
Thus one estimates that
\begin{equation}
    \sigma_G^2 \sim a^{6-4\Delta} \,.
\end{equation}
The signal-to-noise from correlating a single pair of planes on a single lattice is therefore on the order of 
$G(\tau)/\sigma_G \sim (a/\tau)^{2\Delta - 3} \ll 1$.

The gradient flow method \cite{Narayanan:2006rf,Luscher:1984xn} offers an approach to ameliorate the short-distance divergent behavior in these correlation functions.
Rather than evaluating the correlation functions directly on the lattice configuration, one first applies a well-defined procedure to remove UV fluctuations in the fields down to a length scale $\sim \sqrt{8\tf}$, where $\tf$ is the gradient-flow depth.
This reduces the divergent short-distance behavior such that
$\int d \vec{x} \langle \OO^\dagger(\vec{x},0) \OO(\vec{0},0) \rangle \sim (8\tf)^{(3-2\Delta)/2}$.
Physical results require an extrapolation to small $\tf$, which partially counteracts the gain in signal-to-noise.
Nevertheless, it is still necessary to apply some additional kind of noise-reduction technique in order to get a signal in a reasonable amount of computing time.%
\footnote{
An alternative noise-reduction technique is the use of the multilevel algorithm \cite{Luscher:2001up}.
The multilevel method has been successfully applied to correlation functions relevant for transport
\cite{Meyer:2007ic,Meyer:2007dy,Francis:2015daa,Brambilla:2020siz}. However, this technique is only applicable to pure-glue theories; it cannot directly be generalized to the unquenched case. 
Furthermore, it is implemented using Monte Carlo updates of the gauge fields, rather than during the calculation of correlation functions on generated configurations.
Ideas on the implementation of multilevel algorithms including dynamical fermions can be found in Refs.~\cite{Ce:2016idq,Giusti:2017ksp}.
Reference~\cite{GarciaVera:2016dau} presented an approach for combining multilevel and gradient flow techniques.}

This leaves, however, the problem that the signal-to-noise does not improve as one makes $L$ large.
One might have hoped for such improvement, because boxes with larger transverse extent $L$ should be generating more statistically independent samples.
But we do not see such an improvement in our parametric estimates, nor in simulations.
To see why, we look at the role of transverse integrations in the signal and in the noise.  
In \Eq{Gestimate} we see that only $|\vec x - \vec y| \lesssim \tau$, that is, small transverse separations, contribute to the signal.
But \Eq{sigmaestimate} contains $independent$ integrations over $d\vec x$ and $d\vec y$; all values of $\vec x - \vec y$ contribute equally to the noise.
So points with small transverse difference are responsible for signal and noise,
but points with large transverse difference contribute to the noise, but not to the signal.

In this work we propose a blocking technique which eliminates the noise contributions from large transverse separations, and
restores the expected behavior that the signal-to-noise ratio improves as $(L/\tau)^{3/2}$.
The technique is numerically cheap and can be used in conjunction with gradient flow.
In addition, it is applied at the analysis level, not as part of the configuration generation, and it is perfectly compatible with unquenching.

Naturally we are far from the first people to confront this particular problem.
The issue of rapid falloff in the signal but not the noise has been known for a long time \cite{Parisi:1983ae}, and has been discussed and confronted frequently in the recent literature
\cite{Bali:2009dz,Sun:2015enu,Blum:2015gfa,Ce:2016ajy,Shintani:2015vsx,Luscher:2017cjh}.
In particular, Ref.~\cite{Liu:2017man} presented an approach which is in some ways similar to what we argue for here.
We will discuss the relative advantages of the approaches after giving an exposition of what we propose.

\section{Blocking method}
\label{sec:blocking_method}

\begin{figure}[htb] 
\centerline{\includegraphics[trim={60 50 60 80},clip,width=0.5\textwidth]{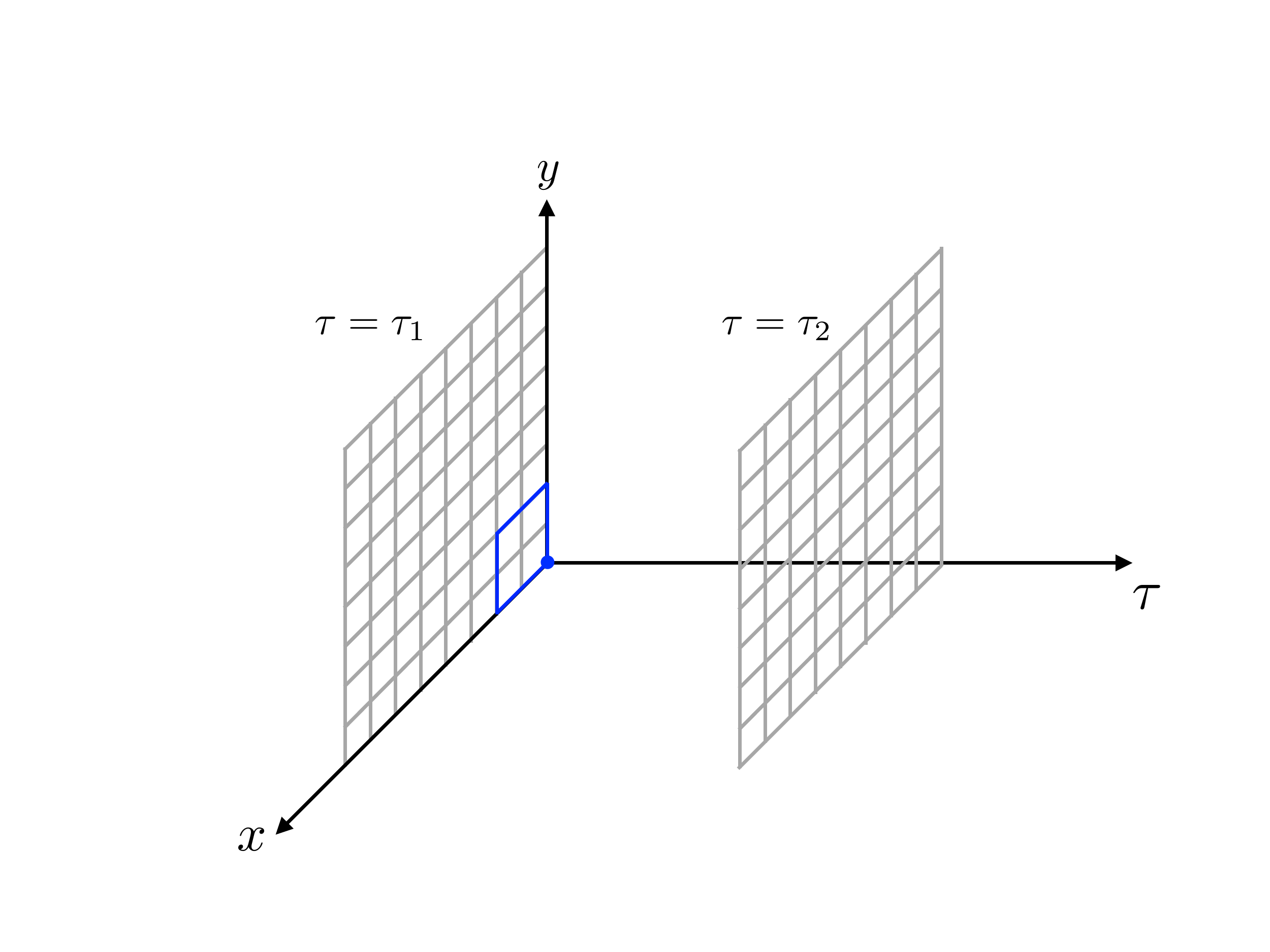}
}
\caption{Illustration of a 3D lattice on which a temporal correlator is measured. Operators at each site of plane $\tau=\tau_1$ are summed and the same is done for $\tau=\tau_2$. Two summed operators will be correlated via \Eq{usual_corr}.
}
\label{sketch_usual_corrs}
\end{figure}

Let us specialize to Hermitian operators $\OO$ and consider the lattice form of the correlation function, where space integrals are replaced by discrete sums:
\begin{align}
G(\tau_1-\tau_2)=\frac{a^3}{V} \sum_{\vec{x}\in V}\OO(\tau_1,\vec{x})\sum_{\vec{y}\in V}\OO(\tau_2,\vec{y}),
\label{usual_corr}
\end{align}
where $V=N_xN_yN_z$ is the spatial volume of the lattice. 
To calculate it one first evaluates the operator on each site on the plane at temporal position $\tau_1$ and also those at $\tau_2$. The two planes are shown as grey squares at $\tau=\tau_1$ and $\tau=\tau_2$ in a simplified 3D sketch in \cref{sketch_usual_corrs}.
Then one calculates all site-to-site correlators of two operators: one operator runs over all sites on plane $\tau=\tau_1$ while the other is fixed to one site on plane $\tau=\tau_2$ and the process is repeated for each site on the second plane. 
The economical and also the most common way to do this is to first compute the sum of the operators on the plane at $\tau=\tau_1$ and then repeat this procedure for all planes. The data on each plane can be reduced to a single number, thus saving memory.

The method mentioned above includes contributions from all possible spatial distances ($s=\sqrt{(\vec{x}-\vec{y})^2}$).
But as we have seen, the signal is dominated by small $s \lesssim \tau_2-\tau_1$.
Therefore, we want to obtain differential information, namely, how the correlation function varies as a function of $s$.
A complete differential measurement involves correlating each $\vec x$ position with each $\vec y$ position.
The numerical cost of this scales as $V^2$, which is prohibitive, so we must seek a numerically less costly alternative, which we now present.

\begin{figure}[htb] 
\centerline{
\includegraphics[trim={60 50 60 80},clip,width=0.5\textwidth]{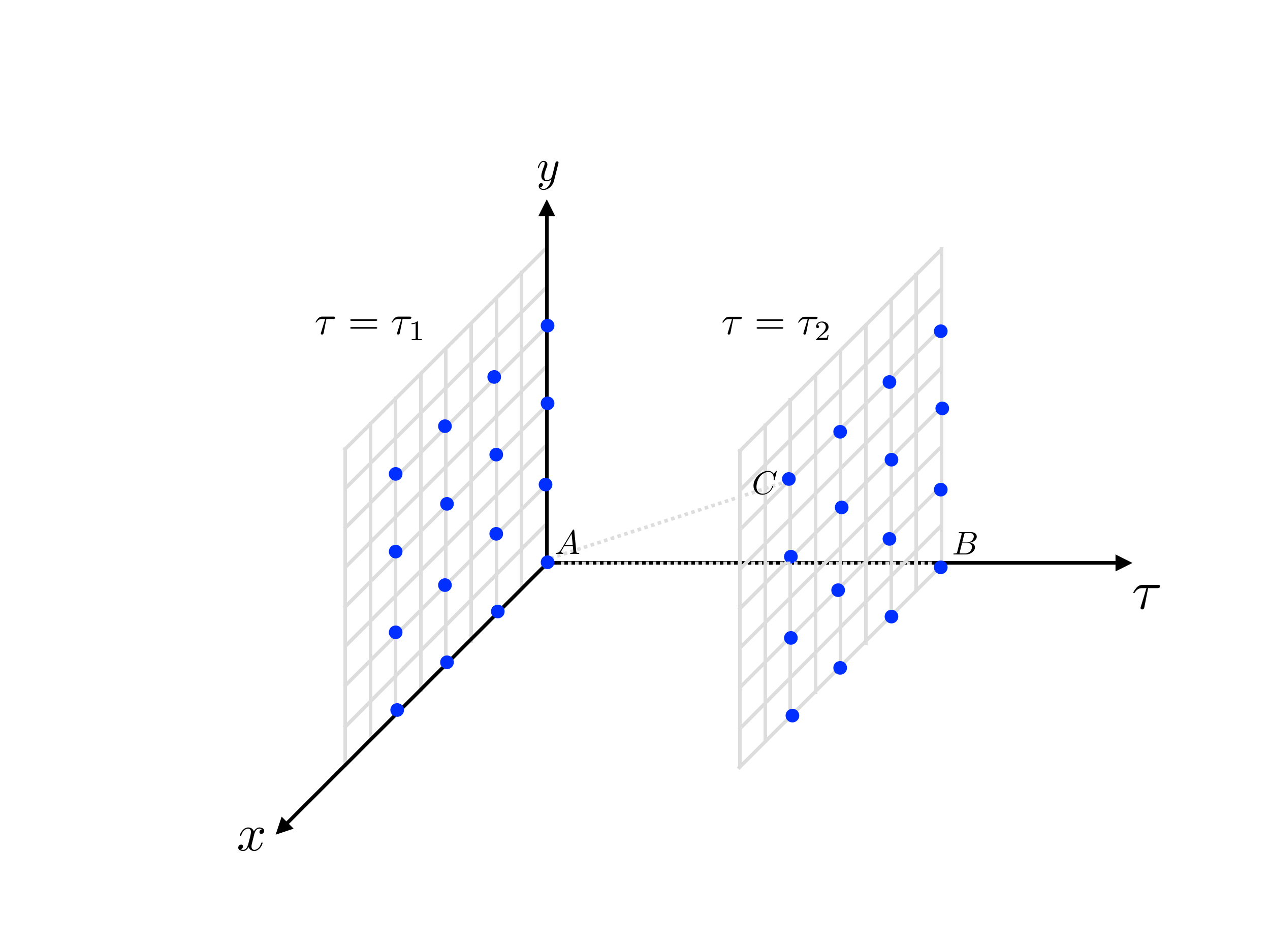}
}
\caption{Illustration of how the blocking method works: each plane is divided into bins. The operators in each bin are summed and saved to the site denoted by the blue dot on the corner.
Then, for each pair of planes, one computes the correlators of all pairs of blue dots, one on each plane.
}
\label{sketch_blocking_corrs}
\end{figure}

\begin{figure*}[bth] 
\centerline{\includegraphics[width=0.5\textwidth]{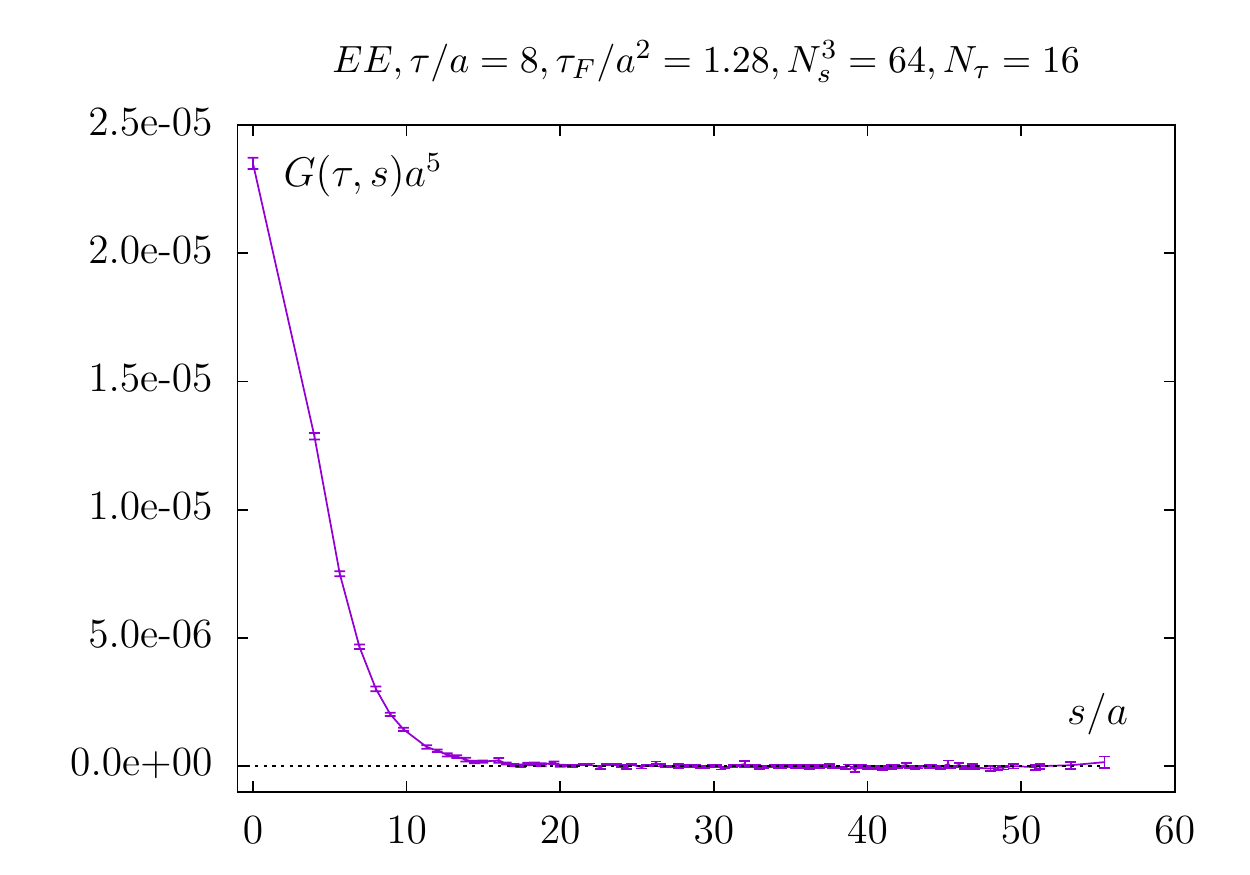}
\includegraphics[width=0.5\textwidth]{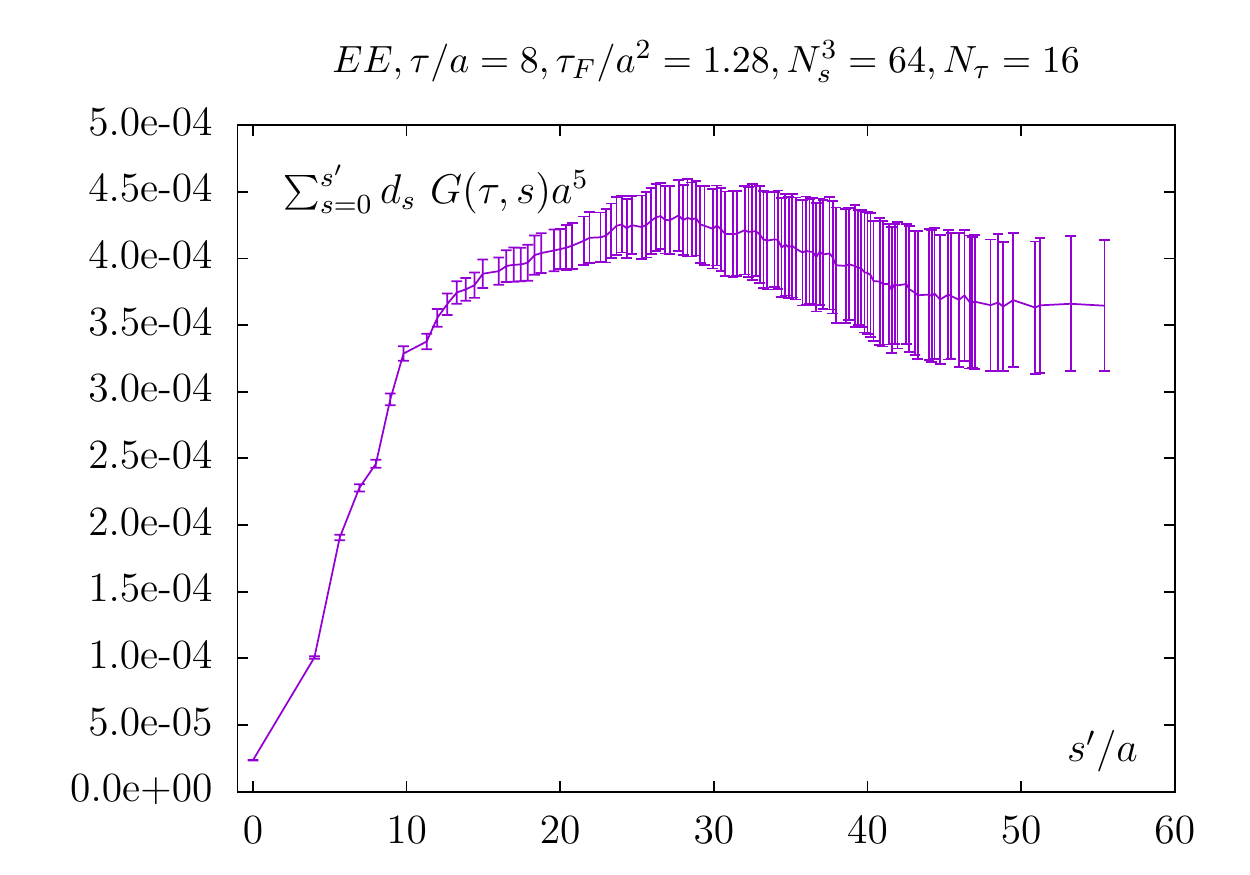}
}
\caption{Left: (bare) energy-momentum tensor correlator as a function of spatial distance $s$ on a $16 \times 64^3$ lattice at temperature $T=1.5\,T_c$ in the bulk channel at temporal separation $\tau/a=8$ and flow depth $\tf/a^2=1.28$.
The correlator is measured using the blocking method on $10\,000$ quenched configurations. 
The bin size is $4\times 4\times 4$.
Right: sum of the correlator (sum of all terms $s \leq s'$) as a function of the maximal spatial distance $s'$.
The last data point (at the largest $s'$) is equal to the correlator calculated in the conventional way (without blocking) using \Eq{usual_corr}.
}
\label{bulk_nodegen_sum}
\end{figure*}

In the blocking method, each plane, for instance the $\tau=\tau_1$ plane in \cref{sketch_usual_corrs}, is split into equal-sized bins.
In each bin, we measure the operator on all sites belonging to this bin and save the sum at one corner.
For instance, the blue square in \cref{sketch_usual_corrs} encloses one bin of size $2\times 2$ and the sum is denoted as a dot at the origin.
The relative position of this dot inside of the bin should be the same for all bins.
Covering the whole lattice with bins leads to \cref{sketch_blocking_corrs}, in which the original lattice has effectively been compressed into a smaller lattice denoted by the blue dots.
On this smaller lattice a calculation of the point-to-point correlators at all possible distances is feasible.
One can see this procedure is nothing but rewriting \Eq{usual_corr} as
\begin{align}
G(\tau)=\frac{a^3}{V} \!\! \sum_{\{v_1\} } \! 
 \left[ \sum_{\vec{m}\in v_1}\OO(\tau_1,\vec{m}) \right]
 \sum_{\{v_2\} } \left[\sum_{\vec{n}\in v_2}\OO(\tau_2,\vec{n}) \right]
\label{blocking_corr}
\end{align}
where $\tau=\tau_1-\tau_2$, $v_1,v_2$ are the individual bins in each plane, and $\vec m \in v_1$ are all points in the bin $v_1$.
We advocate the use of cubic bins, that is, each bin contains $V_b = n_b \times n_b \times n_b$ lattice sites.
Defining
\begin{align}
    \OO_{v_1}(\tau_1) &= \sum_{\vec m \in v_1} \OO(\tau_1,\vec m)
    \nonumber \\
    G_{v_1 v_2}(\tau) & = \langle \OO_{v_1}(\tau_1)
    \OO_{v_2}(\tau_2) \rangle
\end{align}
we can reexpress \Eq{blocking_corr} as
\begin{equation}
    G(\tau) = \frac{a^3}{V} \sum_{v_1,v_2} G_{v_1 v_2}(\tau) \,.
\end{equation}
It is useful to rearrange this sum in terms of the transverse separation between the corner points of the bins $s = |\vec v_1 - \vec v_2|$.
For $s=0$ there are $V/V_b$ contributions (equal to the number of bins).
For any other $s$ value there is an additional degeneracy factor $d_s$.
For instance, for $s=n_b$ we can have
$\vec v_1 - \vec v_2\in \{ [\pm n_b,0,0],[0,\pm n_b,0],[0,0,\pm n_b]\}$
for a total of $d_s=6$ degenerate choices.%
\footnote{Normalizing $s$ to $s^2/n_b^2=0,1,2,3,\ldots$, the $d_s$ with $s^2\leq(L/2n_b)^2$ are the OEIS integer sequence A005875; see https://oeis.org/A005875.}
Introducing the sum of all correlations between bins with separation $s$, normalized to contain $V/V_b$  contributions,
\begin{equation}
\label{Gtaus}
    G(\tau,s) = a^3d_s^{-1} 
    \sum_{v_1,v_2} G_{v_1 v_2}(\tau) \delta(|\vec v_1-\vec v_2|-s)\,,
\end{equation}
we can write the total correlation function as
\begin{equation}
    G(\tau) = \frac{1}{V_b} \sum_{s=0}^{s_{\rm{max}}} d_s G(\tau,s) \,,
\end{equation}
where $s_{\rm{max}}^2=3(\frac{L}{2})^2$.

This representation will be particularly practical in the following.
Specifically, $G(\tau,s)$ should be a smooth function of $s$, which allows us to fit its behavior in a range of $s$ where the signal-to-noise is good, and to use this fit to estimate its behavior at large $s$, where the signal-to-noise is bad.

A similar decomposition can be achieved using Fourier techniques, as described in Ref.~\cite{Liu:2017man}.
The authors show how fast Fourier transform techniques can determine the transverse-separation-by-separation correlator with of order $V \log V$ operations.
(In comparison, our approach requires of order $V^2/n_b^6$ operations;
in practice, either scaling renders the required compute time small compared to the time required to update and gradient-flow gauge configurations, except perhaps on extremely large lattices.)
To get something analogous to \Eq{Gtaus}, 
one could histogram the resulting separation-by-separation correlator into transverse-separation ranges, which allows a transverse-separation-differential analysis of the data which is analogous to what we achieve here.
Therefore we consider the Fourier method of Ref.~\cite{Liu:2017man}, together with some histogramming, to be an alternative to our blocking technique, with very similar advantages.
The main difference between our approach and the approach of Ref.~\cite{Liu:2017man} will be in how we use this differential information.
The next section will show how we extract the integrated correlator,
applying it to two specific physical problems.

\section{Applications of the blocking method}
\label{sec:applications}

Let us illustrate how to take advantage of blocked-correlator information in a real calculation which can be used to study transport phenomena.

Bulk viscosity can be determined from the small-frequency behavior of the spectral function for the squared field strength operator
\begin{equation}
\label{E}
E(x)=\frac{1}{4}F^a_{\rho\sigma}(x)F^a_{\rho\sigma}(x).
\end{equation}
To compute bulk viscosity on the lattice, we first need the zero-momentum correlation function
\begin{equation}
\label{corr_EE}
G(\tau)\equiv \int d^3\vec{x}\ \langle \delta E(0,\vec{0})\delta E(\tau,\vec{x})\rangle
\end{equation}
as a function of $\tau$.
Here $\delta E(x)\equiv E(x)-\langle E(x)\rangle$
is the field strength with its expectation value subtracted off to remove the disconnected contributions.
In our implementation, we construct $E$ using the clover definition of the field strength tensor.
In this work we will focus on determining this correlator with good signal-to-noise.
The issues of correctly normalizing the operator, continuum and zero-flow extrapolating, and extracting the bulk viscosity from $G(\tau)$ are left for a separate study.
We measure this correlator using the blocking method on a $64^3\times 16$ quenched lattice over $10\,000$ configurations under gradient flow, determining errors using the bootstrap method.
The bin size is $4\times 4\times 4$. The lattice setup and gradient flow setup are the same as used in Ref.~\cite{Altenkort:2020axj}. 
We will also revisit the correlation function of topological charge density $q$ which we first addressed in Ref.~\cite{Altenkort:2020axj}, using the same $q$ definition introduced there and the same lattice setup as just described.

We illustrate the EMT correlators calculated using the blocking method in \cref{bulk_nodegen_sum}.
The left panel shows the correlator $G(s,\tau)$ 
as a function of $s$ at a fixed flow time $\tf/a^2=1.28$ and temporal separation $\tau/a=8$, where $a$ is the lattice spacing.
We can see that $G(s,\tau)$ is a relatively smooth function of $s$, and that it falls off  fast such that only the first few data points ($s/a\lesssim 15$) contribute significantly to $G(\tau)$.
At distances $s/a>17$ the correlators cannot be statistically distinguished from zero.
If one sums the correlator over all $s \leq s'$, including the degeneracy factor, one obtains the data shown in the right panel of \cref{bulk_nodegen_sum}.
The data point at the largest $s'$ recovers the result and errors calculated in the usual (nonblocking) way.
It can be seen that at $s'/a \sim 15$ the integrated correlator reaches a plateau but the error size becomes larger and larger as $s'$ increases.
From this it is clear that the bin-to-bin correlators with small $s$ are contributing most of the signal while the large-$s$ ones mainly introduce noise. 
The key idea of the blocking method is to use only the reliable lattice data coming from small $s$ and to estimate the contribution from the long tail by fitting the data that has good signal-to-noise using some theoretically inspired $Ansatz$.

For each $(\tau,\tf)$ pair, we break the $s$ range into three regions based on two cut points, $s_0$ and $\scut$.
The first region, $s < s_0$, is characterized by very high signal-to-noise ratios in the data $G(\tau,s)$.
The point $s_0$ is chosen as the largest $s$ value where the signal-to-noise is better than 10.
This point is very easy to find from the data, and its value is stable across different bootstrap samples.
We will justify this choice $a\ posteriori$, after describing the rest of the fitting procedure.
The middle region $s_0 \leq s \leq \scut$, is characterized by signal-to-noise between 10 and 2.
$\scut$ is determined in a self-consistent way which we will explain soon.
Finally, there is the region $s > \scut$, where the signal-to-noise is very poor.
Our procedure is to perform a fit of the data, and to replace a direct evaluation of $G(\tau)$ with an evaluation which takes into account the fit, as follows:
\begin{align}
    \label{threepart}
    & G(\tau)  = G_{\mathrm{dom}}(\tau) + G_{\mathrm{mid}}(\tau)
    + G_{\mathrm{tail}}(\tau) \,,
    \nonumber \\
     &G_{\mathrm{dom}}(\tau) \equiv  \frac{a^3}{V_b} \sum_{s=0}^{s_0-1}
    d_s G(\tau,s) \,,
    \nonumber \\
     &G_{\mathrm{mid}}(\tau) \equiv  \frac{a^3}{V_b} \sum_{s=s_0}^{\scut} 
    d_s \left( x G_{\mathrm{fit}}(\tau,s) + (1-x) G(\tau,s) \right)\,,
    \nonumber \\
    &G_{\mathrm{tail}}(\tau)  \equiv  \frac{a^3}{V_b} \sum_{s>\scut}
    d_s G_{\mathrm{fit}}(\tau,s) \,.
\end{align}
Here $x=(s-s_0) / (\scut-s_0)$ is the fraction of the way
from $s_0$ to $\scut$; that is, in the first region we purely use the data, in the middle region we vary linearly from purely using the data at $s=s_0$ to purely using the fit at $s=\scut$, and in the final region we purely use the fit.

To perform a fit of the data we need two things: an $Ansatz$, and a data range to use in the fit.
We will return to the $Ansatz$ momentarily.
First we need to emphasize what $s$ range the fit needs to be precise in.
It is not important to find a fit function which describes the whole $s$ range.
As we see in \Eq{threepart}, we only use the fit for $s > s_0$.
We know on physical grounds that $G(\tau,s)$ falls rapidly at large $s$, and this should be reflected in our $Ansatz$.
Therefore, the data range where $G_{\mathrm{fit}}(\tau,s)$ is most
important is the range around $s_0$ and $\scut$.
This fact needs to be reflected both in our choice of $Ansatz$, and in the data range we use to fit the $Ansatz$.
In particular, using data with $s < s_0$ in our fitting procedure is actually not a good idea.
The high signal-to-noise tends to control the fit, but it gives information about the functional form too far away from the region where we need the fit to work.
Therefore, we choose instead to fit the data with $s \geq s_0$.
We could cut off the $s$ range used in the fitting procedure, for instance, at the not-yet-established value $\scut$, but in practice the fit is always dominated by the first few points above $s_0$, even if we use 
$all$ data with $s \geq s_0$.
This is in fact what we do: we fit a physically well-motivated $Ansatz$ to all data with $s \geq s_0$.
We have checked that the fit, $\chi^2$, and errors in the fit parameters are almost unaffected by introducing an upper cutoff on the $s$ range used in the fit.

With a fit in hand, we can then estimate the signal-to-noise ratio as $\mathrm{STN}(s) = G_{\mathrm{fit}}(\tau,s) / \sigma(\tau,s)$, where the noise is determined from the fluctuations in the data and the signal is estimated from the fit.
We choose $\scut$ to be the $s$ value above which this estimated signal-to-noise ratio is always worse than 2.
That is, we fully replace the data with the result of our fit starting where the signal-to-noise is consistently below 2.
We estimate the errors in $G_{\mathrm{dom}}$, $G_{\mathrm{mid}}$ and $G_{\mathrm{tail}}$, whose sum gives the total correlator, using the bootstrap method.

\begin{figure*}[t] 
\centerline{\includegraphics[width=0.5\textwidth]{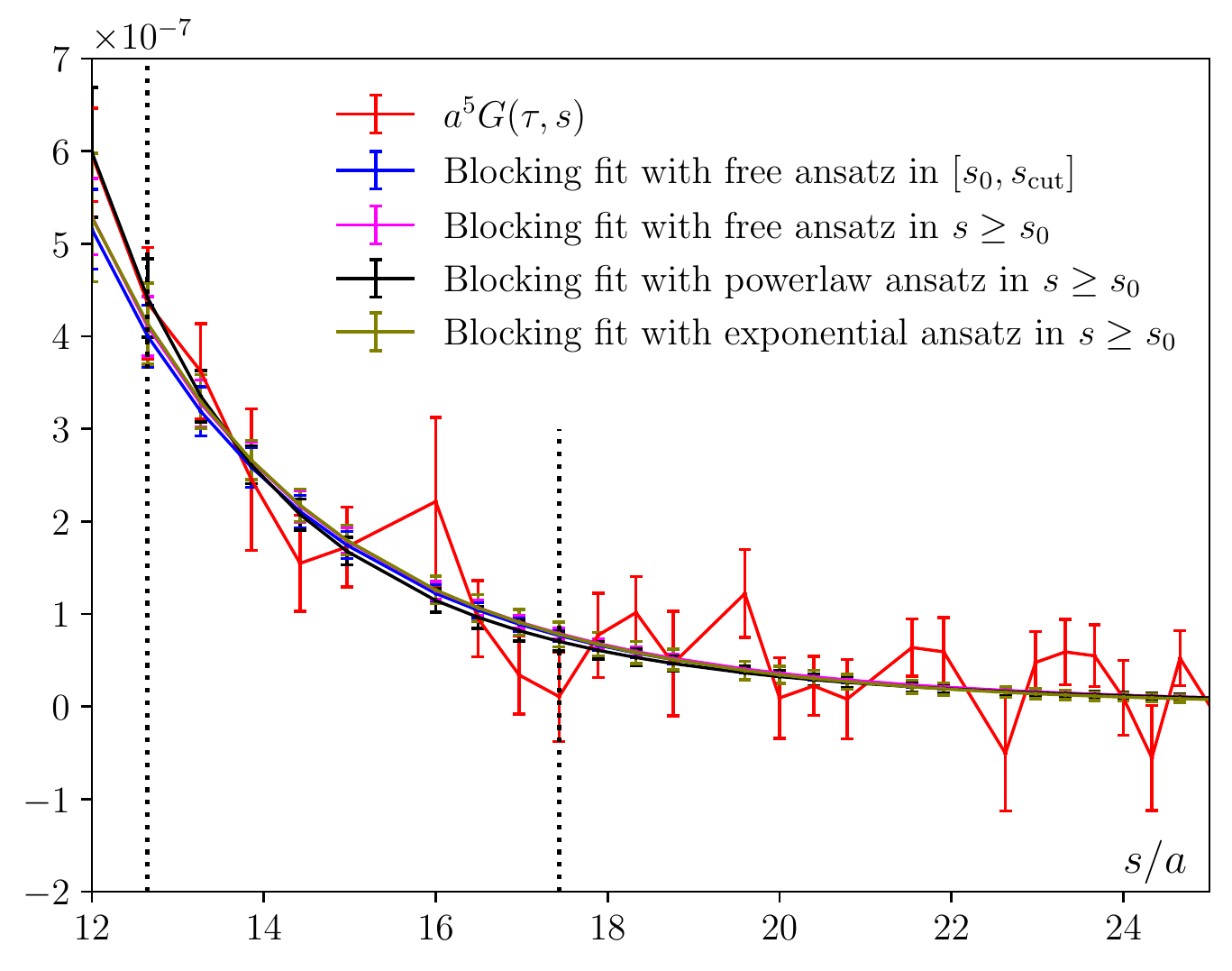}
\includegraphics[width=0.5\textwidth]{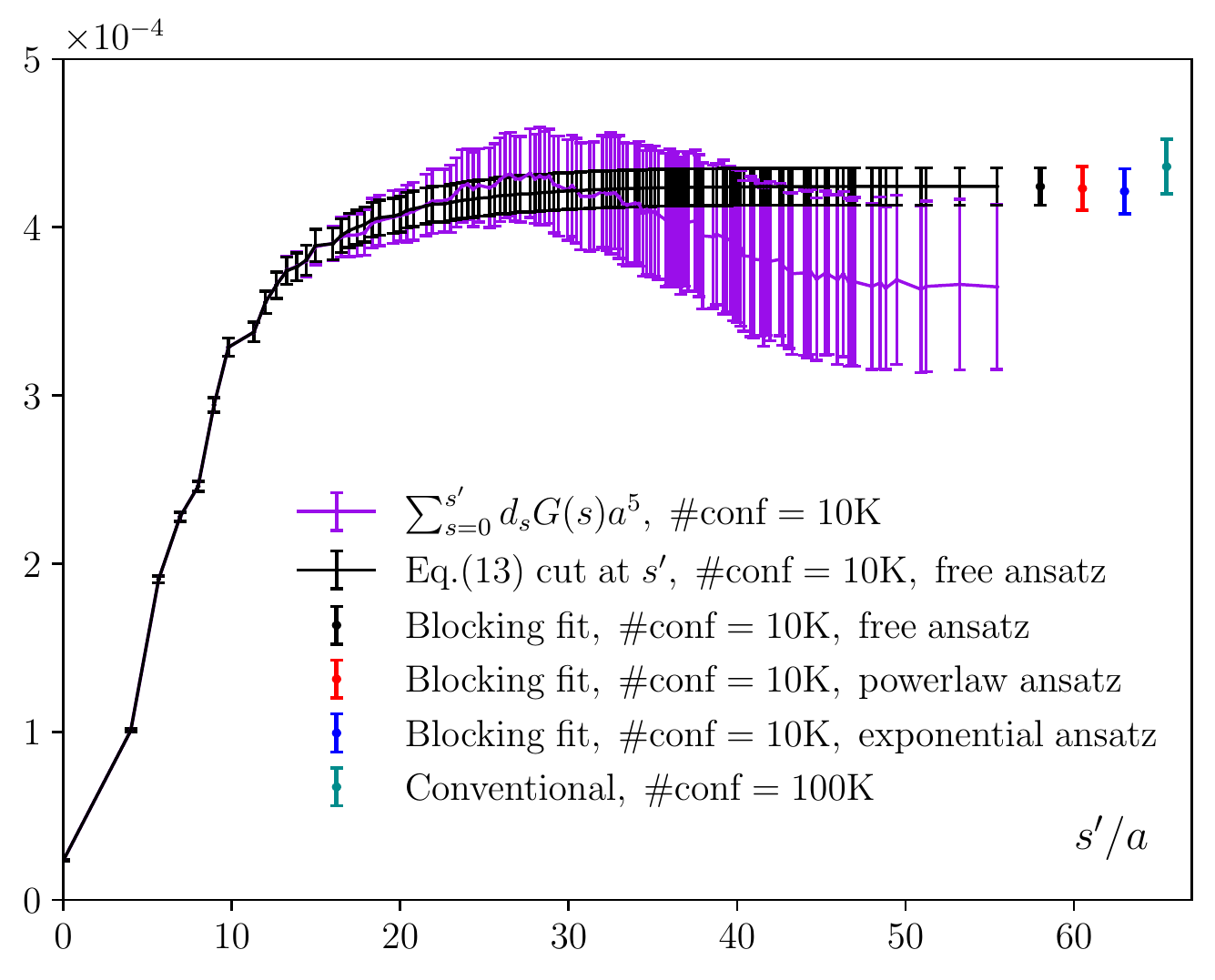}
}
\caption{Left: the same data as in the left panel of \cref{bulk_nodegen_sum}, but zoomed in around the value where the signal-to-noise becomes poor,
together with three fits of the data, representing fitting functions and fit ranges as described in the text.
 Right: the same as the right plot in \cref{bulk_nodegen_sum}, but also showing the result summed over $s \leq s'$ using the mix of data and $Ansatz$ defined in \Eq{threepart}.  Also, the fully summed result from \Eq{threepart} (using power law and free $Ansatz$) and from the conventional approach using 10 times as many configurations are shown (with a slight horizontal offset) as single data points on the far right.
}
\label{bulk_fit_combine}
\end{figure*}

Next we discuss the $Ansatz$($Ans$\"a$tz$) used in the fits.
The choice of $Ansatz$ is clearly dependent on the specifics of the theory under consideration; here we specialize to correlation functions of the action density $E$ and of the topological charge density $q$.

The first $Ansatz$ is a simple power law in $s$,
$a^5G(s,\tau) = A (s/s_{\mathrm{pivot}})^{-B}$, where $A$ and $B$ are fit parameters and $s_{\mathrm{pivot}}$ is the third $s$ value larger than $s_0$.\footnote{$s_{\mathrm{pivot}}$ is introduced to suppress the correlation between $A$ and $B$, which stabilizes the fit.}
The motivation is that a rapidly falling function is generically fit over a narrow range by a power law.
Also, \Eq{Gestimate} implies that the true falloff could be a power law, with an unknown coefficient due to operator anomalous dimensions.%
\footnote{If the falloff is with distance, one should really fit to $(s^2+\tau^2)^{-B/2}$.  Since $\scut$ is almost always significantly larger than $\tau$, this turns out to make very little difference.}

The second $Ansatz$ is based on the leading-order perturbative value for the correlator, accounting for time periodicity, gradient flow, and our blocking procedure.
In vacuum, the leading-order correlator of two field strength tensors is
\begin{align}
    \label{vacFF}
    &\langle F^a_{\mu\nu}(r) F^b_{\alpha\beta}(0) \rangle
    = \frac{g^2 \delta_{ab}}{\pi^2 r^4}
    \left[ \delta_{\mu\alpha} \delta_{\nu\beta} - \delta_{\mu\beta} \delta_{\nu\alpha}
    \vphantom{\frac{2}{r^2}} \right.
    \nonumber \\ & \left.
     - \frac{2}{r^2} \left(
     r_\mu r_\alpha \delta_{\nu\beta} - r_\mu r_\beta \delta_{\nu\alpha} - r_\nu r_\alpha \delta_{\mu\beta}
    + r_\nu r_\beta \delta_{\mu\alpha}
     \right)
    \right].
\end{align}
Applying gradient flow to a depth $\tf$ modifies this expression to
\cite{Eller:2018yje}:
\begin{align}
    \label{vacFFflow}
    &\langle G^a_{\mu\nu}(r) G^b_{\alpha\beta}(0) \rangle_{\tf}
    = \frac{g^2 \delta_{ab}}{\pi^2 r^4}
    \left[ A(r,\tf) \left( \delta_{\mu\alpha} \delta_{\nu\beta} - \delta_{\mu\beta} \delta_{\nu\alpha} \right)
    \vphantom{\frac{2}{r^2}} \right.
    \nonumber \\ & \left.
     + \frac{B(r,\tf)}{r^2} \left(
     r_\mu r_\alpha \delta_{\nu\beta} {-} r_\mu r_\beta \delta_{\nu\alpha} {-} r_\nu r_\alpha \delta_{\mu\beta}
    {+} r_\nu r_\beta \delta_{\mu\alpha}
     \right)
    \right],
    \\ &
    A(r,\tf) = 1 - \left(1+\frac{r^2}{8\tf} \right) e^{-r^2/8\tf}  \,,
    \\ &
    B(r,\tf) = -2 + \left[ 2 - 2 \frac{r^2}{8\tf}
    + \left(\frac{r^2}{8\tf} \right)^2\right]
    e^{-r^2/8\tf} .
\end{align}
Note that this is a continuum, not lattice, expression; but when $\tf /a^2 > 0.5$ the lattice-continuum difference for flowed correlators is small, and the use of a continuum limit at fixed flow depth based only on data which satisfies this criterion should avoid the need to include lattice spacing corrections as well.

\begin{widetext}
Using these expressions, the leading-order $\langle EE \rangle$ correlator at finite $\tf,\tau,|\vec r|$ and with periodic boundaries in the time direction is
\begin{align}
    \langle E(\vec r,\tau) E(0,0)\rangle_{\tf} & \propto
    \sum_{n_1,n_2 \in \mathcal{Z}}
    \frac{A(r_1) A(r_2)}{r_1^4 r_2^4}
    + \frac{A(r_1) B(r_2) + A(r_2) B(r_1)}{2 r_1^4 r_2^4}
    + \frac{B(r_1) B(r_2)}{6 r_1^6 r_2^6}
    \left( 2 (r_1\cdot r_2)^2 + r_1^2 r_2^2 \right),
\end{align}
where $r_1=(\tau+n_1 \beta,\vec r)$ and 
$r_2=(\tau+n_2 \beta,\vec r)$ are the 4-displacement with the temporal displacement shifted by independent integer multiples of the inverse temperature $\beta$.
Similarly, when we compute the correlation function of two topological charge density operators $q = F_{\mu\nu} \tilde F_{\mu\nu}/32\pi^2$ (see below), the leading-order correlation function after flow is
\begin{align}
    \langle q(\vec r,\tau) q(0,0)\rangle_{\tf} & \propto
    - \sum_{n_1,n_2 \in \mathcal{Z}}
    \frac{A(r_1) A(r_2)}{r_1^4 r_2^4}
    + \frac{A(r_1) B(r_2) +A(r_2) B(r_1)}{2 r_1^4 r_2^4}
    + \frac{B(r_1) B(r_2)}{6 r_1^6 r_2^6}
    \left( r_1^2 r_2^2 - (r_1\cdot r_2)^2 \right).
\end{align}

\end{widetext}

\begin{figure*}[thb] 
\centerline{\includegraphics[width=0.48\textwidth]{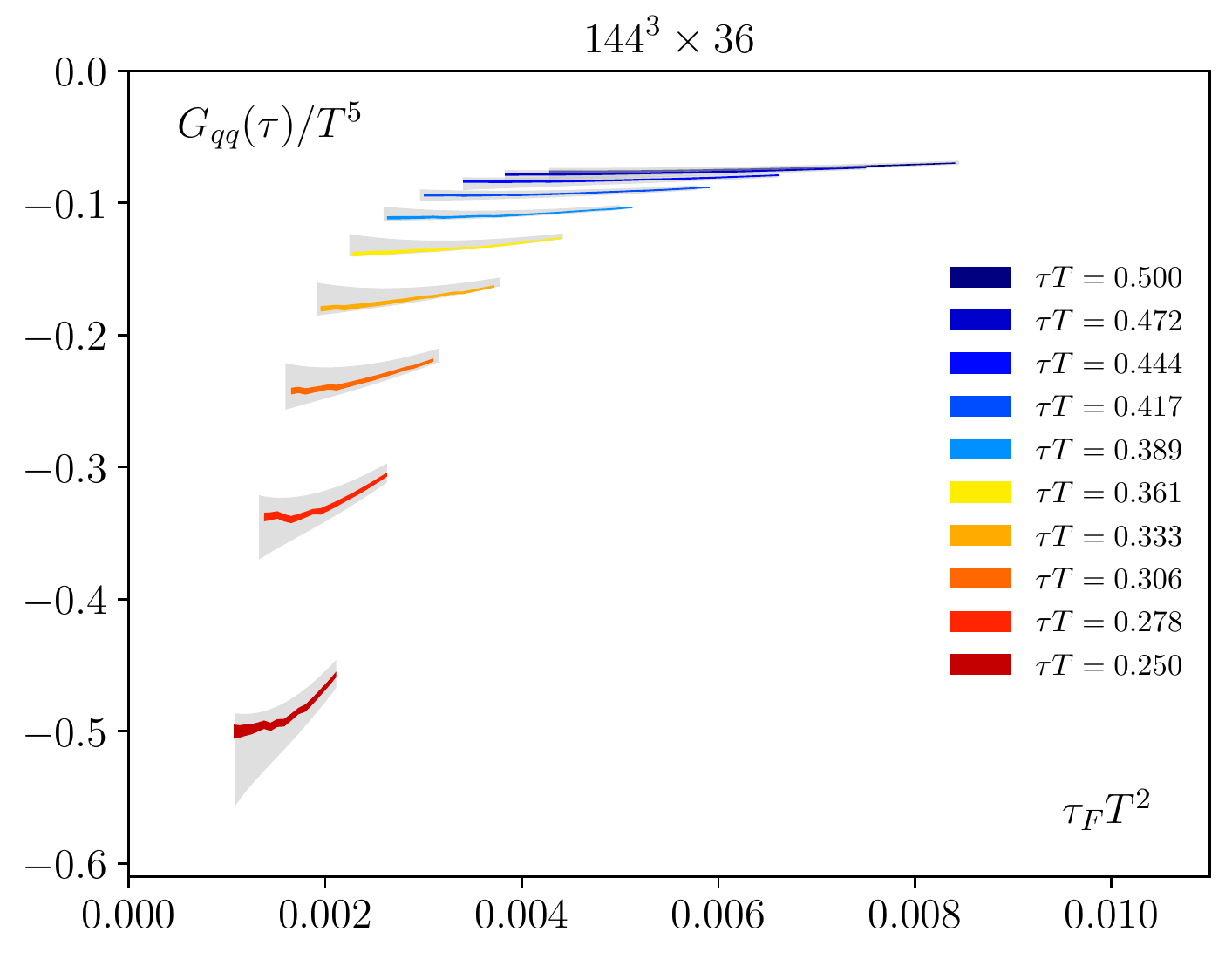}
\includegraphics[width=0.52\textwidth]{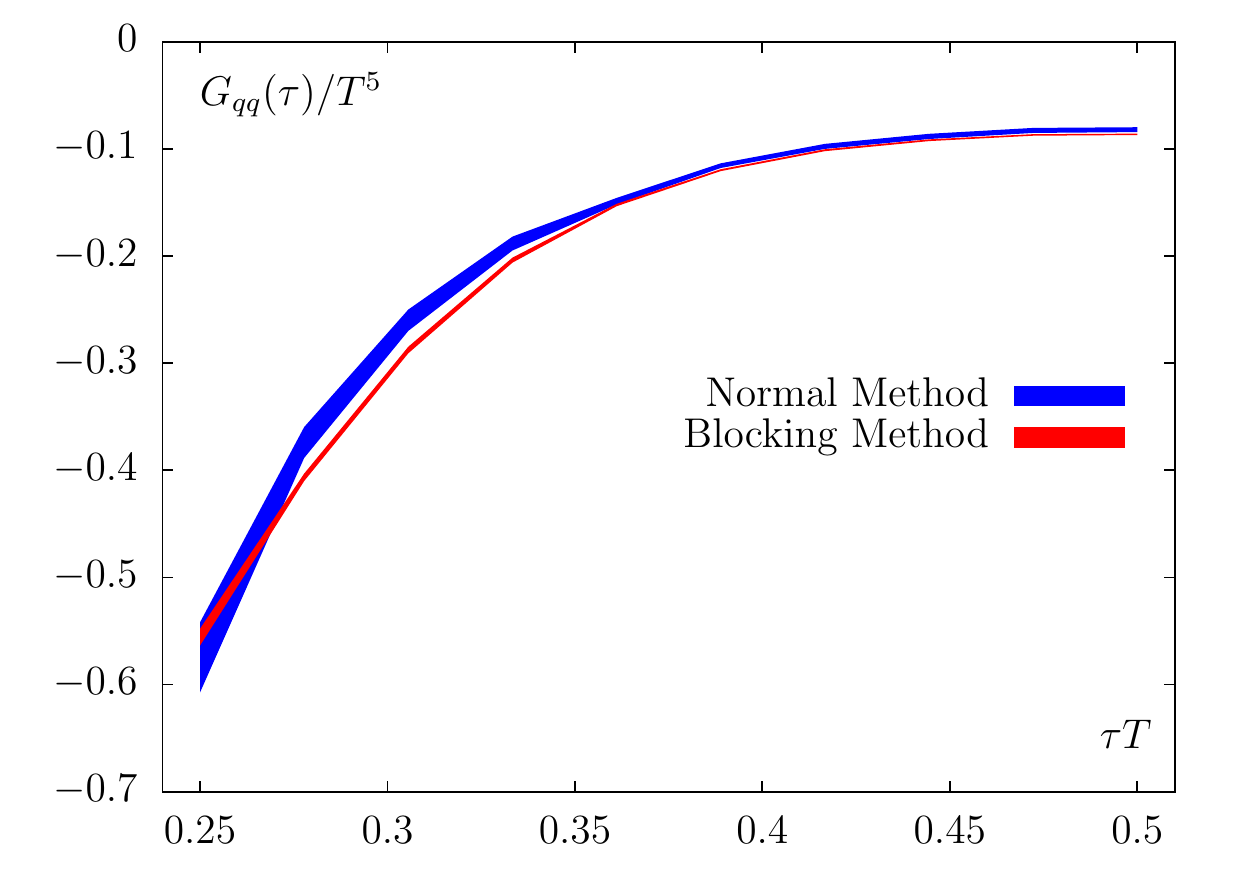}
}
\caption{Left: a comparison of the topological charge density correlators measured in the conventional way on $10\,000$ configurations taken from Ref.~\cite{Altenkort:2020axj} (grey bands) and those from blocking fits (colorful bands) on the same configurations on the $144^3\times 36$ lattice. Right: the same as in the left panel, but after continuum and flow-time-to-zero extrapolation.
}
\label{topo_fit_combine}
\end{figure*}

Our second $Ansatz$ is, for each $s$ value, to integrate these expressions over the relative coordinates in the two bins whose centers are separated by $s$ to determine what the time-periodic, flowed, bin-averaged correlation function would be at leading perturbative order.
We then fit to a single overall normalization.
This fit gives a poor description of the $whole$ correlator $G(s,\tau)$, because it does not get the ratio of the peak to the tail accurately.
However, it gives a reasonable description of the tail shape with a single fitting parameter.

Our third $Ansatz$ will be based on the generic expected long-distance behavior for correlation functions in a nontrivial interacting theory.
Consider a correlator as a function of space separation $s$ along the $z$ axis.
If we think about the $z$ axis as ``Euclidean time,'' then this corresponds to a zero-temperature system in a space with one compact periodic direction $\mathcal{R}^2 \times S^1$.
This theory is expected to have a gapped spectrum, and the large-distance correlation behavior is controlled by the lightest state in the symmetry channel under investigation.
One then expects that the correlation function decays at asymptotically large distances as%
\footnote{%
In $\mathcal{R}^3 \times \mathcal{R}$,
$a^5G \propto (s/s_{\rm{pivot}})^{-3/2} \exp(-B(s-s_{\mathrm{pivot}}))$
(see for instance Ref.~\cite{Luscher:2017cjh}),
but the presence of a small-radius $S^1$ makes the large-distance $s \gg 1/T$ decay behave like that in $\mathcal{R}^2 \times \mathcal{R}$.}
$a^5G(s,\tau)=A(s/s_{\rm{pivot}})^{-1}\exp(-B(s-s_{\rm{pivot}}))$,
where $A,B$ are two fitting parameters.
This $Ansatz$ makes sense if one assumes that $s$ is large enough to suppress the contributions of all higher excited states.

Some readers may be concerned about the $ad$ $hoc$
nature of our fitting functions.
It is important to emphasize three things:
\begin{enumerate}[label=\roman*.]
    \item
The function $G(s,\tau)$ falls very fast with $s$.
    Perturbatively, $G(s,\tau) \propto s^{-8}$.
    We see this rapid falloff explicitly in \cref{bulk_nodegen_sum}, both in the data and in the fits.
    Therefore, it is only important that our fit to $G(s,\tau)$ be reasonable in a fairly narrow range above $s_0$. 
\item
The fit is stable when we introduce an upper $s$ cutoff on the fit range, and it is stable when we adjust (somewhat) the starting point $s_0$ for the fit.
\item
The influence of the fit $Ansatz$ is small, as determined by comparing these three rather different $Ans$\"a$tz$.
\end{enumerate}

To illustrate these points, consider \cref{bulk_fit_combine}.
The left panel shows the same data as in \cref{bulk_nodegen_sum}, but zoomed in around the region where the signal-to-noise becomes poor.
The vertical bars indicate $s_0$ and $\scut$, the red points are the data, and the indicated lines are three fits to the data.
We see that all three fit $Ans$\"a$tz$ give nearly the same results, and the free-$Ansatz$ fit is almost unchanged when we fit to the data in the range $[s_0,\scut]$ rather than all data with $s\geq s_0$.
The fit indicates that the correlation function becomes very small for $s$ a little higher than $\scut$, consistent with the data but without the large error bars.

In the right panel of \cref{bulk_fit_combine} we show a comparison of the correlators measured in the conventional way on $100\,000$ configurations (green data point), fitted correlators based on blocking data on $10\,000$ configurations with the free $Ansatz$ (black points), and the same using the power-law $Ansatz$ (red data point) and exponential-decaying $Ansatz$ (blue point). These are compared to the same partial sums as in \cref{bulk_nodegen_sum}. We can see that if we fit the tail of our data to a proper $Ansatz$, we can reduce the error by a significant factor (in this case by a factor of $\sim 4$). Comparison with a much larger data set shows that this is achieved without corrupting the value; the blocking method gives a result which is consistent with that achieved by the conventional method using 10 times more data.

Now we return to discussing our choice of $s_0,\scut$.
Table~\ref{tab:s0scut} lists the values of $s_0$ and $\scut$ determined at some typical separations $\tau$ and flow times $\tf$.
We see that $s_0$ is primarily determined by the gradient flow depth, and secondarily by $\tau$.
$\scut$ is additionally weakly dependent on our $Ansatz$ choice.
Since $s_0,\scut$ are determined from the data signal-to-noise, they will also change as we increase/decrease the computational resources and hence the amount of fit data.
Specifically, more data means better signal-to-noise which means larger $s_0,\scut$.  Our approach always balances $s_0$ such that the fit is only used where the signal-to-noise has become problematic.

Our choice for $s_0$ was based on the somewhat arbitrary criterion of a signal-to-noise of 10.
We have explored what happens when we either loosen or sharpen this criterion.
If we choose a smaller value for the signal-to-noise, the fit ceases to be very constraining, and we fail to get a good description of the tail.
But what if we start the fit where the signal-to-noise is much higher, say, 50?
We explored this possibility, and our choice, over the full range of $\tau,\tf$ values which we use in our data analysis.
We consistently find that the $\chi^2/\mathrm{d.o.f}$ of our fits, using the data in the range $[s_0,\scut]$, is close to 1 for the criterion we use.
But when we set $s_0$ at the point where the signal-to-noise is 50, we find that the smallest-$\tf$ results$-$where the data is the noisiest$-$produce a poor (large) $\chi^2/\mathrm{d.o.f}$, apparently because the fit relies on data at a small $s$ value where none of our fitting functions are good descriptions.
This $\chi^2$ analysis provides an $a\ posteriori$ justification for our $s_0$ choice: it gives fits which are constraining but which are also internally consistent.

To understand our procedure better, Table~\ref{tab:contribution-each-part} shows a decomposition of the total determined $G(\tau)$ based on the three regions
defined in Eq.(\ref{threepart}), together with the fitted parameters for each $Ansatz$.
It can be seen that the first two $Ans$\"a$tz$ give almost the same contribution for each part.
The exponential $Ansatz$ has a somewhat larger $\scut$,
such that the middle and tail regions contain a different number of points.
But the sum of these two regions is approximately the same in all cases.
The difference between the results using different $Ans$\"a$tz$ is about 10 times smaller than the overall statistical uncertainty.

\begin{table}[h]
    \centering
    \begin{tabular}{ccccccc}                            
    \hline \hline
    $\;N^2_{\sigma}\times N_{\tau}\;$ & $\;n_b\;$ & $\;\tau/a\;$ & $\;\;\tau_F/a^2\;\;$ & $Ansatz$ & $\;s_0/a\;$ & $\;\scut/a\;$
    \tabularnewline
    \hline
    $64^3\times 16$ & 4 & 4 & 1.28 & free & 13.9 & 17.9  \tabularnewline
    $64^3\times 16$ & 4 & 8 & 1.28 & free & 12.6 & 17.4  \tabularnewline
    $64^3\times 16$ & 4 & 8 & 0.605 & free & 8.0 & 12.0  \tabularnewline
    \hline
    $64^3\times 16$ & 4 & 4 & 1.28 & power law & 13.9 & 18.3 \tabularnewline
    $64^3\times 16$ & 4 & 8 & 1.28 & power law & 12.6 & 17.4 \tabularnewline
    $64^3\times 16$ & 4 & 8 & 0.606 & power law & 8.0 & 12.0 \tabularnewline
    \hline
    $64^3\times 16$ & 4 & 4 & 1.28 & $\;$exponential$\;$ & 13.9 & 18.8  \tabularnewline
    $64^3\times 16$ & 4 & 8 & 1.28 & exponential & 12.6 & 17.9 
    \tabularnewline
    $64^3\times 16$ & 4 & 8 & 0.605 & exponential & 8.0 & 12.6   \tabularnewline
    \hline \hline
    \end{tabular}
    \caption{ 
    $s_0$ and $\scut$ values for the correlator defined in \Eq{corr_EE} using our three fitting $Ans$\"a$tz$,  based on $10\,000$ independent configurations at the indicated values of lattice size, block size, and $\tau,\tf$ values.}
    \label{tab:s0scut}
\end{table}

\begin{table*}[htb]
    \centering
    \begin{tabular}{cccccc}                            
    \hline \hline
    $Ansatz$ & $\;G_{\mathrm{dom}}\times 10^6\;$ & $\;G_{\mathrm{mid}}\times 10^6\;$ & $\;G_{\mathrm{tail}}\times 10^6\;$ & $A$ & $B$ \tabularnewline
    \hline
    free & 355.5 (6.6) & 44.6 (11.8) & 24.3 (1.9) & $5.5\times 10^{-5}$ ($4.3\times 10^{-6}$) & $\cdots$ \tabularnewline
    power law & 355.5 (6.6) & 43.7 (11.5) & 23.9 (6.7) & $2.1\times 10^{-7}$ ($1.7\times 10^{-8}$) & 5.7 (0.56) \tabularnewline
    $\;$exponential$\;$ & 355.5 (6.6) & 46.2 (12.1) & 19.3 (6.9) & $\;2.2\times 10^{-7}$  ($1.8\times 10^{-8}$)$\;$ & $\;$0.27 (0.049)$\;$ \tabularnewline
    \hline \hline
    \end{tabular}
    \caption{
    A comparison of the three fitting $Ans$\"a$tz$, showing the fit coefficients and the decomposition of the correlator
    shown in Eq.(\ref{threepart}) for each Ansatz.
    The other values are the same as in each middle line of
    Table~\ref{tab:s0scut}.}
    \label{tab:contribution-each-part}
\end{table*}

An alternative approach, advocated in Ref.~\cite{Liu:2017man}, is to use physical arguments to determine the $s$ value where almost all of the signal has been included, and to discard the data at higher $s$ values.
In \cref{bulk_fit_combine}, this would correspond to using the purple data point at an $s'$ value somewhere above $s'/a=20$.
We see that this approach would be consistent with ours, but with larger errors.

As an application of our technique, we reanalyze the topological charge density correlators which we originally explored in Ref.~\cite{Altenkort:2020axj}.
The correlation function under study is
\begin{align}
G_{qq}(\tau)=\int d^3 \vec{x}\ \langle q(\vec{0},0) q(\vec{x},\tau)\rangle,
\label{gqq}
\end{align}
where the topological charge density is defined as
\begin{align}
q(x)=\frac{g^2}{32\pi^{2}}\epsilon_{\mu\nu\rho\sigma}\textrm{Tr}\left\{ F_{\mu\nu}(x)F_{\rho\sigma}(x)\right\}.
\label{q_def}
\end{align}
Our implementation constructs this operator using an improved field strength tensor $F_{\mu\nu}(x)$; see Ref.~\cite{Altenkort:2020axj} for details.
We repeat the analysis of Ref.~\cite{Altenkort:2020axj} carried out on five lattices$-64^3\times 16$, $80^3\times 20$, $96^3\times 24$, $120^3\times 30$ and $144^3\times 36-$but now applying the blocking method.
The bin size is $4^3, 4^3, 4^3, 6^3, 8^3$ for each lattice respectively.
The number of configurations is $10\,000$ for all lattices.
Other details about the lattice setup and gradient flow setup can be found in Ref.~\cite{Altenkort:2020axj}. 

In \cref{topo_fit_combine} we show a comparison of the correlators measured in the conventional way to those from the blocking method on the same configurations.
Only correlators in the flow time range valid for the $\tf\rightarrow 0$ extrapolation are shown.
In the left panel we take the finest available lattice as an example.
In the right panel we compare the correlators after continuum extrapolation and flow time extrapolation.
From the left panel we can see that the two ways of calculating the correlators give consistent results but with significantly reduced statistical uncertainty at the cost of introducing tiny systematic uncertainty when using the blocking method.
In the right panel a discrepancy between two methods occurs in the range $\tau T\in [0.27, 0.35]$.
This is mostly due to two discrepant points in the original data analysis (not shown) at $\tau T = 0.3, 0.35$ on the $80^3 \times 20$ lattice, which pull the original continuum extrapolation in this region.
The smaller-error results obtained with the blocking method are in better agreement with the other lattice spacings, suggesting that the problem lies in the results obtained without blocking.
Besides these two points, the results using the new approach generally lie within the error bars of the previous determinations.
Using this higher-precision data, we repeat the spectral analysis carried out in Ref.~\cite{Altenkort:2020axj} with the updated correlators, and find that the spectral function $Ans$\"a$tz$ we considered there cannot describe our data well any more.
All the fits have $\chi^2/\mathrm{d.o.f.} >10$.
This indicates that more sophisticated and physically motivated $Ans$\"a$tz$ for the spectral function are needed.
We leave this for future work.

\section{Conclusion}
\label{sec:conclusion}

In this paper we proposed a novel blocking method to improve the signal-to-noise ratio of Euclidean two-point correlators calculated on the lattice.
Taking the bulk channel energy-momentum tensor correlators as an example, we demonstrated a factor of 3$-$7 improvement in the signal-to-noise ratio, with almost no additional cost in numerical effort.
Equivalently, this is a factor 10$-$50 reduction in computational cost to achieve a given precision goal.
We then applied the blocking method to the topological charge density correlators that we studied in a previous publication,
finding that the $Ansatz$ for the spectral function which we previously considered no longer gave a good fit to the data. Our blocking method can be easily implemented on the lattice and used to study various correlators.  It is applied at the analysis level, and does not have to be integrated into the configuration-generating procedure.  There is also no obstacle to using it on unquenched lattices for various bosonic correlators with any physically justified model.

Let us briefly address our choice of bin size.
We chose to use bins somewhat smaller than the largest $\tau$ difference to be considered, in order to get
sufficiently differential information about $G(s)$.
As the bin size is made smaller, the numerical cost to correlate all bins eventually becomes significant.
For the bin sizes considered here, this was not yet a problem.
Also, as we make the bin size smaller, we increase the $relative$ error in each individual bin, which might affect our procedure for choosing $s_0$.
If the bins are chosen smaller than the gradient-flow radius, then data at neighboring $s$ values also becomes correlated, and autocorrelations in $G(\tau,s)$ at nearby $s$ values must be handled carefully.
In the opposite direction, if the bins are too large then we get an insufficiently refined determination of the $s$ dependence of the correlation function.
It might be useful to systematically investigate how bin-size choice affects our procedure, but we leave this for future investigation.

All data from our calculations, presented in the figures of this paper, can be found in Ref.~\cite{datapublication}.

\begin{acknowledgments}
All authors acknowledge support by the Deutsche For\-schungs\-ge\-mein\-schaft
(DFG, German Research Foundation) through the CRC-TR 211 ``Strong-interaction matter under extreme conditions"– project number 315477589 – TRR 211.
The computations in this work were performed on the GPU cluster at Bielefeld University using SIMULATeQCD suite \cite{Altenkort:2021fqk,mazur2021}.
We thank the Bielefeld HPC.NRW team for their support.
\end{acknowledgments}

\bibliography{Bibliography}

\end{document}